\documentclass{article}

\usepackage{arxiv}

\usepackage[utf8]{inputenc} 
\usepackage[T1]{fontenc}    
\usepackage{hyperref}       
\usepackage{url}            
\usepackage{booktabs}       
\usepackage{amsfonts}       
\usepackage{nicefrac}       
\usepackage{microtype}      
\usepackage{lipsum}		
\usepackage{graphicx}
\usepackage{natbib}
\usepackage{doi}

\usepackage{latexsym}
\usepackage{graphicx}
\usepackage{mathptmx}
\usepackage[T1]{fontenc}

\usepackage{natbib}
    \setcitestyle{authoryear,round,aysep={}}

\usepackage{float} 

\usepackage{booktabs} 

\usepackage{multirow} 

\usepackage{longtable}

\usepackage{array}
\newcolumntype{L}[1]{>{\raggedright\let\newline\\\arraybackslash\hspace{0pt}}p{#1}}
\newcolumntype{C}[1]{>{\centering\let\newline\\\arraybackslash\hspace{0pt}}p{#1}}

\usepackage{minted} 

\usepackage{xcolor}

\usepackage{soul} 
\definecolor{light-gray}{gray}{0.95}
\newcommand{\code}[1]{ \sethlcolor{light-gray}{\hl{\texttt{#1}}} } 
\definecolor{light-gray}{gray}{0.95}\newcommand{\codeTitle}[1]{\colorbox{light-gray}{\texttt{#1}}} 
\newcommand{\grayhline}[1]{\arrayrulecolor{lightgray} \cmidrule{2-4} \arrayrulecolor{black}}

\usepackage{booktabs}

\usepackage[table]{xcolor}

\title{MATraM: A Multi-Activity Transport and Mobility Agent-Based Model for Activity Modifications}

\author{ 
    \href{https://orcid.org/0000-0003-2370-6172}{\includegraphics[scale=0.06]{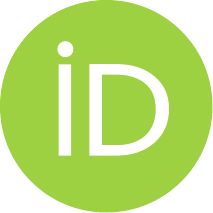}\hspace{1mm}
    Yahya Gamal} \\
	AI for Collective Intelligence Hub\\
	University of Glasgow, UK\\
	\And
    {\includegraphics[scale=0.06]{orcid.pdf}\hspace{1mm}
    Ricardo Colasanti} \\
	School of Geography\\
	University of Leeds, UK\\
    \And
    \href{https://orcid.org/0000-0002-8596-0590}
    {\includegraphics[scale=0.06]{orcid.pdf}\hspace{1mm}
    Gary Polhill} \\
	Information \& Computational Sciences Department\\
	The James Hutton Institute, UK\\
    \And
    \href{https://orcid.org/0009-0007-1195-866X}
    {\includegraphics[scale=0.06]{orcid.pdf}\hspace{1mm}
    Tatsuya Mitomi} \\
	AI for Collective Intelligence Hub\\
	University of Glasgow, UK\\
    \And
    \href{https://orcid.org/0000-0001-9246-3966}
    {\includegraphics[scale=0.06]{orcid.pdf}\hspace{1mm}
    Esra Suel} \\
	Urban Analytics, Department of Geography\\
	University of Zurich, Switzerland\\
    \And
    \href{https://orcid.org/0000-0002-0663-3437}
    {\includegraphics[scale=0.06]{orcid.pdf}\hspace{1mm}
    Alison Heppenstall} \\
	AI for Collective Intelligence Hub\\
	University of Glasgow, UK\\
}

\renewcommand{\shorttitle}{MATraM: A Mobility ABM for Activity Modifications}

\hypersetup{
pdftitle={A template for the arxiv style},
pdfsubject={q-bio.NC, q-bio.QM},
pdfauthor={David S.~Hippocampus, Elias D.~Striatum},
pdfkeywords={First keyword, Second keyword, More},
}

\begin{document}
\maketitle

\begin{abstract}
    This paper introduces the Multi-Activity Transport and Mobility (MATraM) Agent-Based Model (ABM), a novel framework designed to advance activity-based transport modelling by incorporating dynamic activity adaptation. Traditional transport models simulate system performance using varying levels of abstraction, including flow-based, queue-based, and interaction-based mobility representations. While these approaches differ in their treatment of movement and congestion, they typically rely on pre-defined trip patterns that limit responsiveness to changing conditions. In particular, conventional activity-based models generate trips from fixed daily schedules, constraining their ability to capture behavioural flexibility and uncertainty.
    MATraM addresses this limitation by enabling agents to flag modification requests to their planned activities in response to sub-optimal travel conditions, such as increased travel times. By coupling with an activity scheduling and modification framework, the model integrates adaptive decision-making into the generation and execution of daily activity schedules. This allows for a more realistic representation of how individuals adjust their behaviour in response to transport system dynamics, leading to emergent patterns of mobility and congestion.
    The model is presented following the Objectives, Design concepts, and Details (ODD) protocol, outlining its purpose, structure, and implementation. MATraM includes detailed representations of agents, their activity schedules, and the transport network, alongside submodels governing routing, scheduling, and behavioural adaptation. By bridging activity-based modelling with interaction-based mobility simulation, MATraM provides a flexible and extensible platform for exploring transport dynamics under uncertainty. This work contributes to the development of next-generation transport models capable of capturing the complex interplay between individual behaviour and system-level outcomes.

\end{abstract}


\section{Introduction}\label{Introduction}


Transport models are developed to simulate the performance of different components of the transport system. In these models, the mobility of transport system users broadly falls into three categories based on the degree of abstraction: (1) flow-based mobility; (2) queue-based mobility; and (3) interaction-based mobility. 
Flow-based mobility models represent movement as numeric flows between road network nodes \citep[e.g., POLARIS; see][]{Souza2019, POLARIS2025}. It is inspired from fluid dynamics, where at each modelled time step, vehicles are moved across nodes representing flows.
Queue-based mobility models allow the movement to be based on road capacities. If a road link is at maximum capacity, the moving entities (vehicles, buses etc) only enter the road when there is space \citep[e.g, MATSim; see][]{Axhausen2016}. In this case, these entities can remain in the road for the period needed to move through it assuming no congestion.
Interaction-based mobility represents the temporal micro-scale movement on road networks. This allows for simulating the process of reducing speed to avoid collisions. In this case, the focus is the moving entities' behaviours, and congestion becomes an emergent pattern. Individual-based models are particularly relevant in these applications, particularly agent-based models \citep{Heppenstall2021}.

To deploy these models, the trips made in the system are first generated. The trips can be generated broadly through two approaches: (1) supply and demand approach; and (2) activity-based approach. The supply and demand approach estimates the number of trips based on the attraction of spatial zones (demand) and the available transport services given the transport network (supply) \citep{mcnally2000fourstep}. The activity-based approach estimates the number of trips based on the activities that each individual aims to make during a day \citep[e.g.,][]{Pougala2023}. Once trips are defined, the trip modes and routes are selected and optimised to match the initially defined activity schedules. This limits the capacity of the activity-based approaches to account for potential uncertainties in the assigned activities. To address this gap, \citet{Gamal2026} proposes an activity scheduling framework for calibrating mobility models while modifying activities across the population. To support this framework, we introduce the Multi-Activity Transport and Mobility (MATraM) Agent-Based Model (ABM) framework. MATraM expands on activity-based mobility models by allowing activity modification requests in case of sub-optimal trip times. MATraM is developed in NetLogo \citep{Wilensky1999}, and it is available in the following public repository: \href{https://github.com/YahyaGamal/MATraM}{https://github.com/YahyaGamal/MATraM}.

This paper introduces MATraM following the Objectives, Design principles and Details (ODD) protocol \citep{Grimm2006, Grimm2020}. Section \ref{Putpose and patterns} focuses on the purpose and patterns. Section \ref{Entities, state variables and scales} focuses on the entities and state variables. Section \ref{Process overview and scheduling} focuses on the scheduling of events in MATraM. Section \ref{Design concepts} focuses on the design concepts. Section \ref{Initialisation} focuses on the initialisation of the model. Section \ref{Input data} highlights the input data formats and provides sample inputs. Section \ref{Submodels} introduces the submodels deployed during the model runs.


\section{Purpose and patterns} \label{Putpose and patterns}
The purpose of MATraM is to simulate the mobility patterns of individuals following their daily activity schedules through different modes of transport, particularly cars, buses and walking. It also aims to allow individuals to request to modify their daily activities if their trips are not achieved within their tolerance of high trip times.
Daily activity schedules are heterogenous, hence the model generates variable flow patterns across road networks. This yields different patterns of congestion across the system. These congestion patterns contribute to different rates of activity modification requests, alongside the heterogenous tolerance thresholds to long trips. 

\section{Entities, state variables and scales} \label{Entities, state variables and scales}

MATraM includes a set of entities that can be categorised into agents, objects, spatial units and an environment.
First, the model includes three types of agents: (1) cars; (2) pedestrians; and (3) buses. Agents have a set of behaviours and actions that they undertake each time step (see sections \ref{Submodels}).
Second, the model includes two objects: (1) bus templates; and (2) buildings. Both are spatially static and perform no actions.
Third, two types of spatial units exist: (1) nodes; and (2) road links. Spatial objects are vector-based geometries representing street networks, where nodes are connection points between road links. As the model is developed in NetLogo, the spatial units are overlaid on square cells that are used as the model's unit length (e.g., a node of length 10 has a length of 10 square cells). This measuring unit is transformed to miles at initialisation as indicated in section \ref{Initialisation}.
Lastly, the model includes an observer environment \citep{Wilensky1999}, which manages the execution of functions temporally and the global state variables.
A full list of the state variables for the aforementioned entities are shown in Table \ref{tab:1}.

\begin{longtable}{L{0.35cm} L{3cm} L{1cm} L{10.5cm}}
	\toprule
     & State variable & Label & Description\\
	\midrule
    \endhead 
    
    \bottomrule
    \endfoot 

    \endlastfoot 
    
    \multirow{8}{*}{\rotatebox[x=0cm, y=0cm]{90}{Car}}
        & location            & & Current location of the car (origin node on current link)\\ \grayhline{}
        & destination         & & Destination of the car (destination node on current link)\\ \grayhline{}
        & location-time       & & Time step at which the car was at the start node of the current road link\\
        & destination-time    & & Time step at which the car reached the end node of the current road link\\ \grayhline{}
        & top-speed           & & Maximum speed of the car\\ 
        & speed               & & Current speed of the car\\ \grayhline{}
    \multirow{25}{*}{\rotatebox[x=0cm, y=0cm]{90}{Car}}
        & progress            & & Total distance made from the origin node on the current road link\\ \grayhline{}
        & trip                & & The path to ultimate destination is a list of roads\\ \grayhline{}
        & active?             & & If car has not finished all its required trips\\ \grayhline{}
        & at-activity?        & & Whether the car is at a b-destination for an activity or not\\ \grayhline{}
        & my-home             & & Home building of the car\\ \grayhline{}
        & b-destinations      & & The list of all the building destinations (in order)\\ \grayhline{}
        & b-destination       & & The destination building for the car\\ \grayhline{}
        & i-destination       & & The index of the destination building in the b-destinations\\ \grayhline{}
        & activity            & & The list of the types of the activities signposted using integers (in order)\\ \grayhline{}
        & target-times        & \(t\) & The target time steps to reach each destination\\ \grayhline{}
        & times               & & The time steps at which the car reached its destinations\\ \grayhline{}
        & delays              & \(y\) & The delay periods (comparing times to target-times)\\ \grayhline{}
        & expected-periods    & \(p\) & The expected time steps to be taken to reach the destinations\\ \grayhline{}
        & buffer-period       & \(b\) & The number of time steps the car moves earlier to achieve its target-times\\ \grayhline{}
        & periods             & & The list of the periods taken to reach each b-destination\\ \grayhline{}
        & tolerance           & & The tolerance of individuals for long trips\\ \grayhline{}
        & modify?             & & Whether the agent requests an activity modification or not\\ \grayhline{}
        & modify-type         & & Type of activity modification type - can be "cancel" or "reschedule"\\ \grayhline{}
        & initial-step?       & & Whether the car is at the initial trip from home or not\\
    \midrule
    \multirow{26}{*}{\rotatebox[x=0cm, y=0cm]{90}{Pedestrian}}
        & location                      & & Current location of the pedestrian (origin node on current link)\\ \grayhline{}
        & destination                   & & Destination of the pedestrian (destination node on current link)\\ \grayhline{}
        & location-time                 & & Time step at which the pedestrian was at the loaction\\ \grayhline{}
        & destination-time              & & Time step at which the pedestrian reached the destination\\ \grayhline{}
        & top-speed                     & & Maximum speed of the pedestrian\\ \grayhline{}
        & speed                         & & Current speed of the pedestrian\\ \grayhline{}
        & progress                      & & Total distance of the current progress on a road (total distance from origin to reach the destination)\\ \grayhline{}
        & remaining-progress            & & Remaining distance of the current progress on a road (remaining distance from origin to reach the destination)\\ \grayhline{}
        & public-transport?             & & Whether the agent is willing to use public transport or not\\ \grayhline{}
        & trip-legs                     & & The paths or bus services and stop points on each leg of the total trip\\ \grayhline{}
        & trip-legs-modes               & & The mode of transport used in each trip leg ("bus", "walk")\\ \grayhline{}
        & trip-legs-expected-periods    & \(p\) & The expected time steps to be taken to finish each respective trip-leg\\ \grayhline{}
        & trip                          & & The path to achieve the current leg (can be a list of roads or a paired list of a node and a bus template)\\ \grayhline{}
        & i-leg                         & & The index of the current leg\\ \grayhline{}
        & on-bus?                       & & Whether the pedestrian is on a bus or not\\ \grayhline{}
        & waiting?                      & & Whether the pedestrian is waiting for a bus or not\\ \grayhline{}
        & current-bus                   & & The bus which the pedestrian is currently on\\
    \multirow{31}{*}{\rotatebox[x=0cm, y=0cm]{90}{Pedestrian}}
        & passenger-link                & & The passenger link between the pedestrian and the public transport agent (bus)\\ \grayhline{}
        & active?                       & & If pedestrian has not finished all its required trips\\ \grayhline{}
        & at-activity?                  & & Whether the pedestrian is at a b-destination for an activity or not\\ \grayhline{}
        & my-home                       & & Home building of pedestrian\\ \grayhline{}
        & b-destinations                & & The list of all the building destinations (in order)\\ \grayhline{}
        & b-destination                 & & The destination building for the pedestrian\\ \grayhline{}
        & i-destination                 & & The index of the destination building in the b-destinations\\ \grayhline{}
        & activity                      & & The list of the types of the activities signposted using integers (in order)\\ \grayhline{}
        & activity-weight               & \(w\) & The list of the weights given to each activity\\ \grayhline{}
        & target-times                  & & The target time steps to reach each destination\\ \grayhline{}
        & times                         & & The time steps at which the pedestrian reached its destinations\\ \grayhline{}
        & delays                        & \(y\) & The delay periods (comparing times to target-times)\\ \grayhline{}
        & expected-periods              & \(p\) & The expected time steps to be taken to reach the destinations\\ \grayhline{}
        & buffer-period                 & \(b\) & The number of time steps the pedestrian moves earlier than it should to achieve its target-times\\ \grayhline{}
        & periods                       & & The list of the periods taken to reach each b-destination\\ \grayhline{}
        & cancel-periods                & \(c\) & The period thresholds at which an activity modification is requested\\
        & cancel-periods-\%             & \(c^\%\) & The percentage of the expected period that yields a threshold at which an activity modification is requested\\ \grayhline{}
        & tolerance                     & & The tolerance of individuals for long trips\\ \grayhline{}
        & modify?                       & & Whether the agent requests an activity modification or not\\ \grayhline{}
        & modify-type                   & & Type of activity modification type - can be "cancel" or "reschedule"\\ \grayhline{}
        & initial-step?                 & & Whether the pedestrian is at the initial trip from home or not\\
    \midrule
    \multirow{24}{*}{\rotatebox[x=0cm, y=0cm]{90}{Bus}}
        & service             & &  The number of the bus service\\ \grayhline{}
        & location            & &  Current location of the bus (origin node on current link)\\ \grayhline{}
        & destination         & &  Destination of the bus (destination node on current link)\\ \grayhline{}
        & location-time       & &  Time step at which the bus was at the location\\ \grayhline{}
        & destination-time    & &  Time step at which the bus reached the destination\\ \grayhline{}
        & top-speed           & &  Maximum speed of the bus\\ \grayhline{}
        & speed               & &  Current speed of the bus\\ \grayhline{}
        & speed-restriction   & &  Local speed restriction\\ \grayhline{}
        & progress            & &  Total distance of the current progress on a road (total distance from origin to reach the destination)\\ \grayhline{}
        & remaining-progress  & &  Remaining distance of the current progress on a road (remaining distance from origin to reach the destination)\\ \grayhline{}
        & trip                & &  The path to next stop-point destination is a list of roads\\ \grayhline{}
        & active?             & &  If bus has not finished all its required trips (reached all its stop points\\ \grayhline{}
        & at-sp?              & &  Whether the bus is at a stop-point or not\\ \grayhline{}
        & sp-destinations     & &  The list of all the stop-point node destinations (in order)\\ \grayhline{}
        & remaining-sp-destinations & & the list of the remaining sp-destination\\
    \multirow{15}{*}{\rotatebox[x=0cm, y=0cm]{90}{Bus}}
        & sp-destination      & &  The destination stop-point for the bus\\ \grayhline{}
        & i-destination       & &  The index of the destination stop point in the sp-destinations\\ \grayhline{}
        & target-times        & \(t\) &  The target time steps to reach each destination\\ \grayhline{}
        & times               & &  The time steps at which the bus reached its destinations\\ \grayhline{}
        & delays              & \(y\) &  The delay periods (comparing times to target-times)\\ \grayhline{}
        & target-periods      & &  The target periods to make the trips (based on actually incurred bus data)\\ \grayhline{}
        & expected-periods    & \(p\) &  The expected time steps to be taken to reach the destinations\\
        & & & \\
        & periods             & &  The list of the periods taken to reach each b-destination\\ \grayhline{}
        & max-capacity        & &  The maximum capacity of the bus\\ \grayhline{}
        & capacity            & &  The current capacity of the bus (number of passengers)\\ \grayhline{}
        & initial-step?       & &  Whether the bus is at the initial trip from home or not\\
    \midrule
    \multirow{15}{*}{\rotatebox[x=0cm, y=0cm]{90}{Bus template}}
        & service                   & & The number of the bus service\\ \grayhline{}
        & location                  & & Initial location of the bus\\ \grayhline{}
        & active?                   & & Whether the generated bus will not have finished all its trips (always set as true)\\ \grayhline{}
        & at-sp?                    & & Whether the generated bus will be at a stop-point or not (always set as true)\\ \grayhline{}
        & sp-destinations           & & The list of all the node destinations (in order)\\ \grayhline{}
        & remaining-sp-destinations & & The list of all the sp-destinations\\ \grayhline{}
        & sp-destination            & & The first destination building for the bus\\ \grayhline{}
        & i-destination             & & The index of the first destination stop point in the sp-destinations\\ \grayhline{}
        & target-periods            & \(t\) & The expected time steps to be taken to reach the destinations\\ \grayhline{}
        & frequency                 & & The frequency of the bus in time steps\\ \grayhline{}
        & max-capacity              & & The maximum capacity of the bus\\
    \midrule
    \multirow{7}{*}{\rotatebox[x=0cm, y=0cm]{90}{Building}}
        & b-type & & The type of the building\\ \grayhline{}
        & b-xpos & & X-position of the building\\ \grayhline{}
        & b-ypos & & Y-position of the building\\ \grayhline{}
        & b-id   & & ID of the building\\ \grayhline{}
        & nearest-node   & & The nearest node to building\\
    \midrule
    \multirow{15}{*}{\rotatebox[x=0cm, y=0cm]{90}{Node}}
        & n-id   & & ID of the node\\ \grayhline{}
        & n-xpos & & X-position of the node\\ \grayhline{}
        & n-ypos & & Y-position of the node\\ \grayhline{}
        & sp-id  & & Stop-point ID of the node (-1 if not a stop point)\\ \grayhline{}
        & buses-id        & & The service number of the buses passing through the node (empty list if not a stop point)\\ \grayhline{}
        & buses-freq      & & The frequency of the serive (empty list if not a stop point) [minutes]\\ \grayhline{}
        & buses-start     & & Whether the buses starts from the node or not\\ \grayhline{}
        & buses-period    & & The period in time steps until the next bus arrives (empty list if not a stop point)\\ \grayhline{}
        & next-SPs        & & The list of the next stop points for each bus service (the indices refer to the the buses-id in the respective index)\\
    \multirow{1}{*}{\rotatebox[x=0cm, y=0cm]{90}{Node}}
        & buses-schedule  & & The list of the expected time steps at which the bus will reach the stop point (the indices refer to the buses-id in the respective index)\\
    \midrule
    \multirow{16}{*}{\rotatebox[x=0cm, y=0cm]{90}{Road}}
        & r-id                              & & ID of road\\ \grayhline{}
        & distance\_to\_destination         & & Distance to the destination node (in miles)\\ \grayhline{}
        & distance\_to\_destination\_bus    & & Distance to the destination node that can be done by a bus (high dummy value if the bus does not pass) \\ \grayhline{}
        & speed\_limit                     & & Speed limit of the road (in (miles/hour) / 100)\\ \grayhline{}
        & current\_speeds                   & & Current speeds on the road (any element cannot be higher than the speed\_limit)\\ \grayhline{}
        & current\_period                   & \(r\) & Last period taken to make the trip\\ \grayhline{}
        & current\_period\_non\_car         & & Last period taken to make the trip by anything other than a car (shortest period of pedestrians and buses is allocated)\\ \grayhline{}
        & mean\_period\_walk                & & Period taken to a make a trip by a pedestrian walking at the mean input speed\\ \grayhline{}
        & buses?                            & & Whether a bus passes through the road link or not\\
    \midrule
    \multirow{35}{*}{\rotatebox[x=0cm, y=0cm]{90}{Global}}
        & tick-time-scale & & The time in seconds each time step represents\\ \grayhline{}
        & reference-road-id-1 & & The node ID of the first end point of the road link to be used to calculate the model-to-mile-factor\\ \grayhline{}
        & reference-road-id-2 & & The node ID of the second end point of the road link to be used to calculate the model-to-mile-factor\\ \grayhline{}
        & reference-road-miles & & The length of the reference road in miles\\ \grayhline{}
        & synthetic-population? & & Whether the model uses an input synthetic population or not (if not, the model generates the population through the parameters below)\\ \grayhline{}
        & number-of-cars & & The number of cars initialised in the system\\ \grayhline{}
        & number-of-pedestrians & & The number of pedestrians initialised in the system\\ \grayhline{}
        & max-n-activities & & The maximum number of activities an agent targets to achieve per day\\ \grayhline{}
        & return-home? & & Whether pedestrians and cars add home as their last activity or not\\ \grayhline{}
        & max-activity-duration & & The maximum duration of an activity\\
        & max-buffer-period & & The maximum period a pedestrian or car moves earlier than required to reach their next destination on-time\\ \grayhline{}
        & max-home-duration-cars & & The maximum time steps a car agent stays at home before starting its first trip\\ \grayhline{}
        & min-home-duration-cars & & The minimum time steps a car agent stays at home before starting its first trip\\ \grayhline{}
        & max-home-duration-pedestrians & & The maximum duration a pedestrian agent stays at home before starting its first trip\\ \grayhline{}
        & min-home-duration-pedestrians & & The minimum duration a pedestrian agent stays at home before starting its first trip\\ \grayhline{}
        & pedestrians-public-transport & & Whether pedestrians use buses or not\\ \grayhline{}
        & public-transport-search-radius & & The geospatial distance which pedestrians search for stop-points (bus stops) when considering to use buses\\
    \multirow{28}{*}{\rotatebox[x=0cm, y=0cm]{90}{Global}}
        & search-radius-increase-increment & & The increment of increasing the geospatial search radius for public transport in case no stop points are found\\ \grayhline{}
        & fastest-car & & The maximum top speed of cars\\ \grayhline{}
        & slowest-car & & The minimum top speed of cars\\ \grayhline{}
        & fastest-pedestrian & & The maximum top speed of pedestrians\\ \grayhline{}
        & slowest-pedestrian & & The minimum top speed of pedestrians\\ \grayhline{}
        & fastest-bus & & The maximum top speed of buses\\ \grayhline{}
        & slowest-bus & & The minimum top speed of buses\\ \grayhline{}
        & threshold-input-type & & Whether the time thresholds are empirical or generated by the model (if not empirical, the following global variables are used)\\ \grayhline{}
        & mean-tolerance & & The mean tolerance threshold to long trip times\\ \grayhline{}
        & min-delay-threshold & & The minimum delay time steps at which which an activity modification is requested\\ \grayhline{}
        & period-threshold-1 & & The percentage of the min-delay-threshold at which a modification for the activity labelled 1 is requested\\ \grayhline{}
        & period-threshold-2 & & The percentage of the min-delay-threshold at which a modification for the activity labelled 2 is requested\\ \grayhline{}
        & period-threshold-3 & & The percentage of the min-delay-threshold at which a modification for the activity labelled 3 is requested\\
        & & & \\ \grayhline{}
        & period-threshold-4 & & The percentage of the min-delay-threshold at which a modification for the activity labelled 4 is requested\\ \grayhline{}
        & period-threshold-std & & The standard deviation in the percentage-thresholds generated across agents\\

	\bottomrule		
    \caption{Global variables and state variables for cars, pedestrians, buses, bus templates, buildings, nodes and roads}
	\label{tab:1}	
\end{longtable}

\section{Process overview and scheduling} \label{Process overview and scheduling}

At initialisation, the observer applies six steps involving executing a set of submodels.  A more detailed description of all the submodels is available in sections \ref{Initialisation} and \ref{Submodels}.
First, the observed executes the \code{load-network} submodel. This generates the nodes and links representing the road network through which cars and pedestrians will move.
Second, the observed executes the \code{load-buildings} submodel. This involves generating the building objects and assigning their IDs, types and positions. It also involves assigning their closest node.
Third, the observed executes the \code{load-buses} submodel. This generates the bus templates and assigns their frequencies and routes as per a set of input parameters, which can be informed from empirical data.
Fourth, the observed initialises the bus parameters in the networks. It assigns dummy data to the parameter \code{distance\_to\_destination\_bus} in road links through which no buses pass.
Fifth, the observer executes the \code{cars-init} and \code{pedestrians-init} submodels. This generates a number of agents based on the global parameters \code{number-of-cars} and \code{number-of-agents}. If the model is using empirical data, this submodel is skipped. Instead, the observer generates the population from input data. The submodel handling this process is an additional module that can be added based on the data format used.

During the runs, the observer executes four key steps involving three major submodels.
First, the observer requests all the cars and buses to apply the \code{vehicles-move} submodel. This initialises the car trips when needed from the vehicles origin to the activities' destination. It initialises the bus trips at the correct times from their stop-points. It also manages the movement of cars and vehicles during their trips across the road network while avoiding collisions.
Second, the observer requests from all the pedestrians to apply the \code{pedestrians-move} submodel. This initiates the pedestrian trips from their origins to the next activity destination when needed. It assures pedestrians who consider public transport compare alternatives and decide on their route. It also moves pedestrians currently moving to a stop-point or a final destination across the network at their \code{top-speed}.
Third, the observer applies the \code{buses-start-SPs} submodel. This creates the buses at the time step when they should start their route from their initial stop point.
Lastly, the observer monitors whether all the cars and pedestrians have finished their full daily activities or not. If so, the observer terminates the model run and signals its completion.

\section{Design concepts} \label{Design concepts}
\subsection{Basic principles} \label{Basic principles}
The model builds on three key modelling concepts: (1) activity-based modelling; (2) optimal route finding; and (3) mobility modelling. 
First, activity-based modelling is applied during the assignment of activity schedules. This involves generating sequences of activities that require movement from an origin to a destination in space through a mode of transport.
Second, for route finding, the model deploys shortest weighted path algorithms to optimise the trip times. The follows an assumptions that individuals are aiming to minimise their disutility by minimising their trip times. It should be noted that this principle can be modified and expanded by substituting the route finding algorithm in the \code{vehicles-move} and \code{pedestrians-move} submodels.
Third, mobility modelling is applied to simulate the movement of vehicles space as they aim to complete their trips within time restrictions. MATraM follows a detailed vehicle interaction simulation approach. It simulates vehicle behaviours including speeding down to avoid collisions, rather than using road capacities or queue systems (see section \ref{Introduction}).

\subsection{Emergence} \label{Emergence}
The model shows emergence in two aspects: (1) congestion patterns; and (2) trip time distributions. Congestion patterns emerge from the individual vehicle's behaviours as each tries to avoid collisions \citep{Manley2010}. For instance, a singular slow vehicle can lead to periodic congestions on road networks. Trip time bimodal distributions emerge as an outcome in case of random allocation of trip times, origins and destinations in cities. Initial investigations indicate that this is due to the spatial distribution of potential destinations in central city regions, alongside the emergent congestion in central areas. Overall, this yield a high number of moderately long trips going to central areas (initial distribution peak) and increases the trip times of longer trips that pass through the central areas (second distribution peak).

\subsection{Adaptation} \label{Adaptation}
Agents adapt to a set of stimuli to achieve their objectives.
First, cars adapt to other vehicles ahead on their road link. If the car's \code{speed} leads to a collision in the next time step, the car decreases stops and decreases its \code{speed} to zero. This directly achieves the objective of avoiding collisions.
Cars also identify the quickest routes to reach their activity destination while considering congestion. In doing so, they update their \code{expected-periods} which may imply starting their trip earlier to make up for expected congestions. This is in alignment with the two objectives: make the trip in the shortest time; and reach the activity destination on-time or early.

Second, pedestrians adapt their mode of transport choice based on the available alternatives. Pedestrians consider their surrounding stop-points and all the bus alternatives;  they use the \code{search-radius} to dictate the region in which stop-points are considered. If no stop points are found, they adapt by increasing the search region by a \code{search-radius-increase-increment}. An alternative bus route that yields the quickest trip is then selected. Further, they compare the quickest bus trip to a normal walk trip, and select the quickest alternative. These adaptations align with the objective to make the trip in the shortest time.
Pedestrians also adapt their start time to bus schedules -- in case of using buses. They access the \code{buses-schedule} of the stop-point from which they will take the bus. They select the latest bus that will arrive to their activity at the required \code{target-times}. Individuals adapt their trip start time to ensure they catch the preferred bus. This aligns with the objective to reach the activities on-time or early.

Third, buses adapt their moving time from stop-points based on the \code{target-times}. If a bus reaches a stop point earlier than required in the \code{target-times}, the bus waits until the required arrival time. This achieves the objective that buses aim to be maintain the stop-point \code{buses-schedule}, including avoiding leaving earlier that stated.

\subsection{Objectives} \label{Objectives} 
The model includes three key objectives, relevant to different types of agents: (1) minimise trip times (cars and pedestrians); (2) arrive on-time or early to activities (cars and pedestrians); (3) maintain stop-point schedules (buses) and (4) avoid collisions (cars and buses).

First, minimising trip times is a direct objective that is achieved by considering alternative routes and modes of transport where possible (see section \ref{Adaptation}). In all cases, agents consider an objective measure of expected trip times and select the minimum value as the best alternative. This follows a utility maximisation concept if long trip times are framed as a disutility (a negative utility value) -- it should be noted that the agents do not calculate a utility value when making the decisions.

Second, arrival on-time or early is achieved by adjusting the trip start time to the \code{expected-periods} of the respective trip. In doing so, agents consider starting the trip each time step, generating an alternative per time step. The selected alternative is temporally the first one satisfying the condition in equation \ref{eq:1}.
\begin{equation} \label{eq:1}
    T + p_{i,j\rightarrow j+1|T} + b_{i,j+1} \geq t_{i,j+1}
\end{equation}
where \(T\) is the current time step,
\(p_{i,j\rightarrow j+1|T}\) is the expected period incurred by an agent \(i\) to move from the activity \(j\) to the activity \(j+1\) at \(T\),
\(t_{i,j+1}\) is the target arrival time step to the activity \(j+1\) and 
\(b_{i,j+1}\) is the \code{buffer-period} for arriving earlier than the actual activity start time.


\subsection{Prediction} \label{Prediction}
First, prediction is explicitly applied to support the decision of when to start a trip. Cars and pedestrians select and update their optimum routes, and accordingly predict the \code{expected-periods} of their trips. These predictions consider the current congestion patterns for cars and buses, and accordingly can vary across different time steps. Based on that prediction, cars and pedestrians move at time periods that are expected to assure they reach their activity destinations at the \code{target-times}.

Second, prediction is explicitly used to avoid collisions. Cars and buses predict collisions given their current speed trajectory and their surrounding vehicles. Based on that prediction, they stop to avoid colliding with ahead vehicles.

Lastly, prediction is also implicitly applied when buses wait to maintain the \code{bus-schedule} of stop-points. Buses implicitly predict that when they move on-schedule from a stop-point, they will reach the next stop-point at on-time. This may deviate due to low or high congestion rates. This is partly corrected in subsequent stop-points where possible by repeating the wait behaviour.

\subsection{Sensing} \label{Sensing}
Sensing represents the agents' awareness of the positions and state variables of their surrounding entities.

First, cars and pedestrians sense the road network congestion patterns. Cars achieve this by gaining access to the \code{current\_period} state variable of the roads. They find the best route in the system that minimises the trip time (i.e., minimises the total \code{current\_period} if the route).
Pedestrians access the \code{current\_period\_non\_car} and \code{mean\_period\_walk} road state variables. These are used to find the bus and walk trips yielding the shortest trip time to the destination (i.e., lowest total \code{current\_period\_non\_car} route for trips including buses and lowest \code{mean\_period\_walk} route for walk trips).

Second, cars and buses sense the road link length as they move on it. They access the \code{distance\_to\_destination} road state variable, which represents the road length, and compare it to their distance from the road link nodes. This is used to identify when the car or bus reach the end of the current road link.

Third, cars and buses sense the position of other cars and buses ahead on the same road network. This information is used to avoid collision by stopping.

\subsection{Interaction} \label{Interaction}

Cars and buses directly interact with each other as they move on street networks. Each vehicle senses others on the road network and stops if the trajectory given its \code{speed} will lead to a collision. This interaction at an agent level yields macro-scale congestion patterns.

Pedestrians directly interact with buses by taking up a unit of \code{capacity} when they use a bus. This is a mediated interaction between pedestrians as they consume the limited bus \code{capacity} resource. When the bus reaches maximum capacity, pedestrians at stop points are unable to board the bus. This has an impact on their trip time.

\subsection{Stochasticity} \label{Stochasticity}
Stochasticity is only applied in MATraM during the initialisation phase if no input synthetic population is provided. In that case, the model randomly assigns cars and pedestrians to residential buildings (\code{b-type = "residential"}). For each car and pedestrian, it randomly generates a set of parameters. 
First a series of activities (\code{activity}) with a sequence of random arrival times (\code{target-times}) and destinations (\code{b-destinations}) are generated. The minimum number of activities is one activity, and the maximum is based on the input parameter \code{max-n-activities}. 
Second, a randomly generated \code{buffer-period} is assigned to represent variance in the time an agent moves earlier to reach its activity destination. The buffer period ranges from zero to the input \code{max-buffer-period}.
Third, a random \code{tolerance} to long trips parameter is generated. This assures a heterogenous representation of agents that may request an activity modification at a different threshold.

It must be noted that some of the aforementioned stochastically generated parameters are bypassed based on the initialisation input. 
First, if a synthetic population is provided as an input, the stochastic initialisation of activity sequences, target arrival times and destinations is bypassed. This is because a synthetic population is expected to include information on the \code{activities}, \code{target-times} and \code{b-destinations}.
Second, the \code{tolerance} parameter is not stochastically generated if its empirical inputs are provided. This can be attached as a parameter in the synthetic population.

\subsection{Collectives} \label{Collectives}

Collectives are represented in the model during the commute of pedestrians in buses. Pedestrians create  spatial links with the buses they are currently using. Accordingly, they move in the same direction and at the same speed of the bus. This creates a collective of pedestrians in each bus on the road network.

\subsection{Observation} \label{Observation}

These observations are selected in alignment with the aim to support an activity scheduler framework including activity modifications. The activity modification requests in MATraM are an input to the activity scheduler framework. These requests are dependent on the trip times and delays in arrival to activities. According, MATraM focuses on three key observation metrics: (1) trip time; (2) delay period; and (3) activity modifications.
First, mean trip time considers the incurred trip times of all the trips made in the system until a given time step. It is visualised as a time series graph for meant trip times each time step. A histogram of trip times is also generated to show potential emergent distributions.
Second, delay time (\code{delays}) is the number of time steps between the actual arrival \code{times} and the \code{target-times} of each activity. It is visualised as a histogram to identify the distributions of \code{delays} and potentially link them to the activity modification requests.
Third, activity modifications indicate the number of instances a car or a vehicle requested an activity modification due to high \code{delays} that exceed their \code{tolerance} for long trips. This is generated as a time series of the total number of activity modifications until each respective time step.

\section{Initialisation} \label{Initialisation}
\begin{figure}[h]
    \centering
    \includegraphics[width=0.5\linewidth]{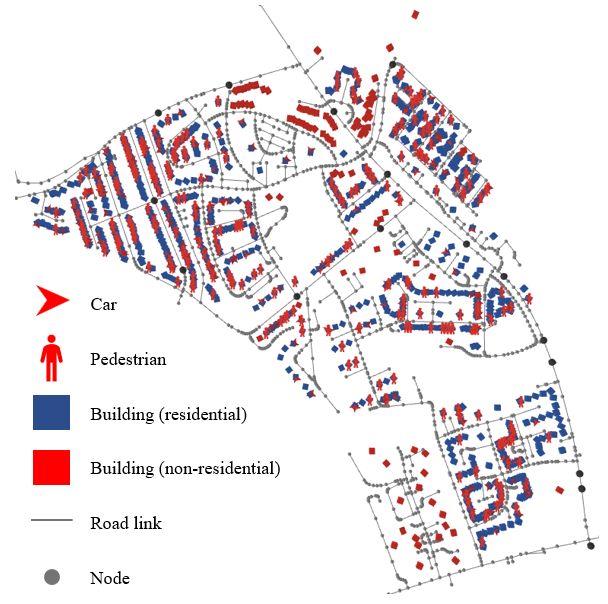}
    \caption{Sample initial state for Tillydrone, Aberdeen, UK}
    \label{fig:init}
\end{figure}

The initialisation in MATraM is designed to allow for transferability across different contexts. It follows three key steps: (1) generate the spatial context; (2) generate the buses; and (3) generate the cars and pedestrians. A set of initialisation submodels are applied sequentially as described in sections \ref{load-buildings}-\ref{pedestrians-init} -- the sequence of submodels during initialisation is also briefly described in section \ref{Process overview and scheduling}. A sample initial state for Tillydrone, Aberdeen, UK is shown in Figure \ref{fig:init}.

\subsection{Submodel: \codeTitle{load-buildings}} \label{load-buildings}
The \code{load-buildings} submodel initialises building objects. It uses input data including the x and y positions of residential and non-residential building (see section \ref{Input data}). Based on that, the spatial locations and types of buildings are assigned (\code{b-type}). The submodel also identifies the closest node to each building. It identifies the node with the shortest Euclidean distance from the building, and it assigns it as the \code{nearest-node}.

\subsection{Submodel: \codeTitle{load-network}} \label{load-network}
The \code{load-network} submodel initialises the node objects and road links. It uses input data including the x and y positions of each node, the IDs of the connected nodes through road links and whether the node is a bus stop point or not (see section \ref{Input data}). 
First, the submodel generates nodes and assign their spatial location. It also initialises its bus related state variables as empty lists, namely \code{buses-id}, \code{buses-freq}, \code{buses-start}, \code{buses-period}, \code{buses-schedule}, \code{buses-schedule} and \code{next-SPs}. These state variable lists are later assigned values in the \code{load-buses} submodel.
Second, the submodel creates undirected road links between the nodes with linked IDs in the input data. It calculates the length of each link and assigns the value to the \code{distance\_to\_destination} state variable. This involves finding the distance between the nodes and transforming it from the model units to miles. Then, it assigns each road link an ID (\code{r-id}) and a maximum speed (\code{speed\_limit}) in miles/hour. Based on the \code{speed\_limit} and the \code{distance\_to\_destination}, the submodel assigns the \code{current\_period} of the road network. At initialisation, this period reflects the time taken by a car moving at the maximum speed to move through the road network. It is later updated during the runs to reflect potential congestion patterns. The submodel also initialises the \code{current\_speeds} state variable as an empty list. This is assigned values as vehicles move across the road links during the runs, and it is used to support the route choice process in the \code{vehicles-move} submodel.

\subsection{Submodel: \codeTitle{load-buses}} \label{load-buses}
The \code{load-buses} submodel initialises the bus templates. It uses input data including the service number, the sequence of stop points, the time period taken to move from one stop point to the next and the bus frequency (see section \ref{Input data}).
The submodel initialises a bus template object and locates it spatially at its initial stop point. It also draws from the input data the state parameters \code{sp-destinations}, \code{remaining-sp-destinations}, \code{target-periods} and \code{frequency}. It also initialises a set of parameters with standard values as shown in Table \ref{tab:2}.

\begin{table}[h]
    \centering
    \begin{tabular}{l l}
    	\toprule
        State variable & value\\
    	\midrule
        active?                   & False\\
        at-sp?                    & True\\
        sp-destination            & None\\
        i-destination             & 1\\
        max-capacity              & 60\\
    	\bottomrule		
    	\label{tab:2}	
    \end{tabular}
    \caption{State variables of bus templates at initialisation}
\end{table}

\subsection{Submodel: \codeTitle{cars-init}} \label{cars-init}
The \code{cars-init} submodels generate the car agents. This defaults to a stochastic generation of car parameters unless a synthetic population is provided. It must be noted that the generation process of cars based on synthetic populations is not included in this model as data input formats can vary across different contexts. Section \ref{Input data} outlines the key parameters required in a synthetic population for MATraM.

First, the submodel creates a number of cars as per the \code{input-number-of-cars}. It assigns each car to a random residential building. It assures the car is labelled as active (\code{active?}), at an activity (\code{at-activity?}) and an the initial step (\code{initial-step?}). It then generates a set of stochastic parameters, including the activity sequences (\code{activity}), based on their respective input variables as shown in Table \ref{tab:3}. The rest of the parameters are initialised as empty lists, and they are addressed in the next submodel steps.

\begin{table}[h]
    \centering
    \begin{tabular}{l l l l}
    	\toprule
        State variable & Relevant input & Distribution & Notes\\
    	\midrule
        activity       & max-n-activities          & Random & No activity repetition \\
        buffer-period  & max-buffer-period         & Random & \\
        tolerance      & mean-tolerance            & Normal & \\
        top-speed      & fastest-car, slowest-car  & Random & \\
        speed          & fastest-car, slowest-car  & Random & \\
    	\bottomrule		
    	\label{tab:3}	
    \end{tabular}
    \caption{Stochastic state variables of cars assigned in the first \code{cars-init} submodel step}
\end{table}

Second, the submodel iterates through each generated activity and assigns a random building as their destination -- this generates the list \code{b-destinations}. It then identifies the route yielding the lowest total \code{distance\_to\_destination} from the car's first activity (home) to the second activity. Accordingly, it assigns a list of links to the \code{trip} state variable of the car. The submodel then deploys the \code{report-expected-period} and \code{report-target-time} sub-submodels per activity. Both sub-submodels identify the shortest routes, based on the road links' \code{distance\_to\_destination}, from an activity to its subsequent one. The \code{report-expected-period} sub-submodel returns an expected time period to reach an activity assuming a vacant system, and this value is assigned to the \code{expected-periods} state variable. The \code{report-target-time} reports a target time step to reach an activity. This is based on the reported expected period and a randomly generated period of time to spend at each activity as shown in equation \ref{eq:2}.
\begin{equation} \label{eq:2}
    t_{j+1} = t_{j} + p_{j\rightarrow j+1} + p_{j+1}
\end{equation}
where \(t_{j+1}\) and \(t_{j}\) are the target times to reach the destination and \(j + 1\) \(j\) given that \(t_j = 0\) for the first activity,
\(p_{j\rightarrow j+1}\) is the time period needed to move from activity \(j\) to activity \(j+1\)
and \(p_{j+1}\) is a randomly generated period to stay in activity \(j+1\) between 0 and the \code{max-activity-duration}.

Third, the submodel then addresses a set of activity related state variables. These variables are lists whose indices are aligned with the indices in the \code{activity} state variable. Table \ref{tab:4} shows the generated values for an element in each of these lists, where each list has the same length as the \code{activity} list state variable.

\begin{table}[h]
    \centering
    \begin{tabular}{l l}
    	\toprule
        State variable & Value of an element in the list \\
    	\midrule
        cancel-periods-\% & Normal(\(\mu\)=period-threshold-n, \(\sigma\)=period-threshold-std) \\
        modify? & False \\
        modify-type & None \\
    	\bottomrule		
    	\label{tab:4}	
    \end{tabular}
    \caption{List state variables of cars addressed in the last step of the \code{cars-init} submodel}
\end{table}

Lastly, the submodel calculates the actual cancel period per activity (equation \ref{eq:3}). It then calculates a normalised activity weight per activity (equation \ref{eq:4}). These values are assigned as lists to the parameters \code{cancel-periods} and \code{activity-weight}, and they provide further information on the thresholds of requesting an activity modification.
\begin{equation}\label{eq:3}
    c_j = c^\%_j.p_j
\end{equation}
\begin{equation}\label{eq:4}
    w_j = 1 - \frac{c_j}{max_{j'\in J}(c_{j'})}
\end{equation}
where \(c_j\) is the cancel-period for activity \(j\),
\(c^\%_j\) is the \code{cancel-period-\%} for activity \(j\),
\(p_j\) is the period spent at activity \(j\),
\(w_j\) is the activity weight and
\(J\) is the set of activities planned by the agent

\subsection{Submodel: \codeTitle{pedestrians-init}} \label{pedestrians-init}

The \code{pedestrians-init} submodel generates the pedestrian agents following a similar logic to the \code{cars-init} submodel. It defaults to a stochastic generation of pedestrians unless a synthetic population is given as an input. The synthetic population pedestrians generation process is not developed in MATraM as data formats vary across different applications. For further details on the required information in a synthetic population for MATraM, see section \ref{Input data}.

First, the submodel creates a number of pedestrians based on the \code{input-number-of-pedestrians}. Similar to cars, it assigns each pedestrian a random residential building as its home. It labels the individual as active (\code{active?}), at the initial time step (\code{initial-step?}) and at an activity (\code{at-activity?}. It then generates a set of stochastic parameters as shown in Table \ref{tab:5}. The state variables of \code{passenger-link} and \code{current-bus} are initialised as \code{None}, and the rest of the parameters are initialised as empty lists.

\begin{table}[h]
    \centering
    \begin{tabular}{l l l l}
    	\toprule
        State variable & Relevant input & Distribution & Notes\\
    	\midrule
        activity       & max-n-activities          & Random & No activity repetition \\
        buffer-period  & max-buffer-period         & Random & \\
        tolerance      & mean-tolerance            & Normal & \\
        top-speed      & fastest-pedestrian, slowest-pedestrian  & Random & \\
        speed          & fastest-pedestrian, slowest-pedestrian  & Random & \\
    	\bottomrule		
    	\label{tab:5}	
    \end{tabular}
    \caption{Stochastic state variables of pedestrians assigned in the first \code{pedestrians-init} submodel step}
\end{table}

Second, the submodel iterates through the \code{activity} state parameter, assigns each activity a building and accordingly generates the \code{b-destinations} list. The shortest route from home to the first activity is then identified assuming pedestrians do no use public transport. This can be later modified during the runs in the \code{pedestrians-move} submodel if a bus trip is deemed more optimum than a walk trip (see section \ref{pedestrians-move}). The submodel then deploys the \code{report-expected-period-pedestrian} and \code{report-target-time-pedestrians} sub-submodels. Both sub-submodels assume pedestrians do not use public transport, and accordingly find the shortest walk routes between each activity and its subsequent one based on the \code{distance\_to\_destination}. The \code{report-expected-period-pedestrian} provides the expected time period to reach each activity assuming no use of public transport. These walk periods represent the longest time a pedestrian takes to reach an activity, as public transport is only used if it yields shorter times during the runs. Further, unlike cars, the pedestrians' trip times remain the same in a vacant or a congested system during the runs. Accordingly, the reported walk periods at initialisation assure the generation of achievable \code{target-times} in the \code{report-target-time-pedestrians} sub-submodel. The calculation of \code{target-times} is applied similar to the \code{cars-init} submodel as shown in equation \ref{eq:2}.

Third, the submodel addresses the \code{cancel-periods-\%}, \code{modify?} and \code{modify-type} state variables. These state variables are lists whose indices reflect the indices in the \code{activity} state variable. Each list element is assigned values similar to the \code{cars-init} submodel as previously shown in Table \ref{tab:4}.

Lastly, the submodel calculates the cancel periods and activity weights for each activity similar to the \code{cars-init} submodel (equations \ref{eq:3} and \ref{eq:4}). The calculated values are assigned to the \code{cancel-periods} and \code{activity-weight} variables.

\section{Input data} \label{Input data}
The model requires input data related to four entities in the model: (1) buildings; (2) nodes; (3) road links; and (4) buses. The data is provided in table format, with columns including necessary information to initialise each entity. The required information for each entity are described hereafter. It must be noted that the table input files must be named as mentioned in this section, and the order of the columns must match the sample table data shown for each entity.

\subsection{Buildings input data}
Buildings require spatial information on their x- and y- positions in space. This spatial information must be normalised from 0 to 1 so that the \code{load-buildings} submodel can correctly assign the building locations. To clarify, for instance, if the model space dimensions are 10 x 10 spatial units, then the an x-position of 0.6 translates to 6 spatial model units in the x direction.

To define the building typology, the x and y positions are provided in two separate tables. The \code{house\_list.csv} table defines homes, and the \code{building\_list.csv} defines non-residential buildings. Table \ref{tab:6} shows a sample input data for either houses or buildings.

\begin{table}[h]
    \centering
    \begin{tabular}{lcc}
        \toprule
        id & xpos & ypos \\
        \midrule
        h\_0 & 0.508 & 0.748 \\
        h\_1 & 0.408 & 0.755 \\
        h\_2 & 0.393 & 0.796 \\
        h\_3 & 0.422 & 0.718 \\
        h\_4 & 0.352 & 0.833 \\
        h\_5 & 0.755 & 0.792 \\
        \bottomrule
    \end{tabular}
    \caption{Sample buildings input data}
    \label{tab:6}
\end{table}

\subsection{Nodes input data} \label{Nodes data}
Nodes require spatial information on their ID,  x and y positions and whether they are stop points or not. First, the node IDs can be provided in any string or numeric format, and they are referred to in the road links input data (section \ref{Road links data}). 
Second, similar to buildings, the spatial information must be normalised from 0 to 1. 
Third, the stop points are indicated through providing an ID where a negative ID indicates the node is not a stop point and positive ID indicates otherwise. The positive ID is later used to indicate the stop points each bus service visits (section \ref{Buses data}). This information is provided in a \code{nodes\_list.csv} table to the model. A sample of the tabular data is provided in Table \ref{tab:7}.

\begin{table}[h]
\centering
\begin{tabular}{lllc}
\toprule
id & xpos & ypos & stoppoint \\
\midrule
n\_id0 & 0 & 0.74 & -1 \\
n\_id1 & 0.012 & 0.701 & -1 \\
n\_id2 & 0.013 & 0.698 & -1 \\
n\_id3 & 0.029 & 0.696 & -1 \\
n\_id4 & 0.051 & 0.696 & -1 \\
n\_id5 & 0.064 & 0.701 & -1 \\
n\_id6 & 0.093 & 0.716 & 11 \\
\bottomrule
\end{tabular}
\caption{Sample nodes input data}
\label{tab:7}
\end{table}

\subsection{Road links input data} \label{Road links data}

Road links require information on their IDs, the IDs of their connecting nodes and their maximum speed in miles per hour.
First, similar to nodes, the IDs can be any string or numeric values.
Second, the connecting nodes IDs refer to the IDs of the nodes provided in the nodes input data (section \ref{Nodes data}). The node IDs are provided as start and end points in the input tabular data.
Third, the maximum speed must be provided as a string in the following format: "<digits> mph". If the maximum speed is provided as a "None" string, the road is assumed to have no speed cap.
Table \ref{tab:8} shows a sample tabular data provided in the \code{link\_list.csv} to the model.

\begin{table}[h]
    \centering
    \begin{tabular}{llll}
        \toprule
        id & start & end & maxspeed \\
        \midrule
        l\_0 & n\_id0 & n\_id1 & 20 mph \\
        l\_1 & n\_id1 & n\_id2 & 20 mph \\
        l\_2 & n\_id2 & n\_id3 & 20 mph \\
        l\_3 & n\_id3 & n\_id4 & 20 mph \\
        l\_4 & n\_id4 & n\_id5 & 20 mph \\
        l\_5 & n\_id5 & n\_id6 & 30 mph \\
        l\_6 & n\_id6 & n\_id7 & 20 mph \\
        \bottomrule
    \end{tabular}
    \caption{Sample road links input data}
    \label{tab:8}
\end{table}

\subsection{Buses input data} \label{Buses data}

Buses require information on (1) their service number, (2) the sequence of the stop point IDs they stop at, (3) the period taken to move from one stop point to its subsequent one, (4) the service frequency at each stop point and (5) whether the stop point is where the bus service starts or not. Unlike the previously mentioned input data, the buses tabular data does not refer to a bus per entry (row). Instead, each entry refer to a stop point visited by a specific bus -- hence one bus service is expected to have multiple entries.

First, the service number can be a numeric or string value.
Second, the sequence of stop points refer to stop point IDs indicated in the nodes input data (see section \ref{Nodes data}). Each row entry indicates one stop point for the  service number with the given ID in that row. Accordingly, the row entries with the same service number dictate the sequence of the stop points for that service number.
Third, the time period needed to reach the stop point from the previous one is provided as a numeric value in minutes. If a stop point is the first one, its time period must be zero.
Fourth, the frequency is provided as numeric value in minutes, and it represents that time between the arrival of two buses with the same service number at each stop point.
Fifth, the indication of whether the stop point is where the service starts is indicated through a binary value; 1 indicates the service starts at this stop point and 0 otherwise.

The data is provided to the model in a file named \code{trips\_list.csv}. For clarity, Table \ref{tab:9} shows a sample input data including two services.

\begin{table}[h]
    \centering
    \begin{tabular}{lllll}
        \toprule
        service & sp-id & time & frequency & start \\
        \midrule
        19 & 1197 & 0 & 5 & 1 \\
        19 & 781 & 0.5 & 5 & 0 \\
        19 & 557 & 1 & 5 & 0 \\
        19 & 592 & 1 & 5 & 0 \\
        111 & 682 & 0 & 10 & 1 \\
        111 & 214 & 1 & 10 & 0 \\
        111 & 716 & 1 & 10 & 0 \\
        111 & 38 & 0.5 & 10 & 0 \\
        \bottomrule
    \end{tabular}
    \caption{Service schedule data}
    \label{tab:9}
\end{table}

\section{Submodels} \label{Submodels}

As stated in the process overview and scheduling (section \ref{Process overview and scheduling}), MATraM includes two key submodels during the runs: (1) \code{vehicles-move}; and (2) \code{pedestrians-move}. These submodels manage the behaviours of cars, buses and pedestrians. A detailed description of each submodel and its respective sub-submodels is provided hereafter.

\subsection{Submodel: \codeTitle{vehicles-move}}
The \code{vehicles-move} submodel manages both cars and buses. For cars, if a car is active (\code{active? = True}), it calls the \code{cars-travel} sub-submodel (section \ref{cars-travel}). For buses, it calls the \code{buses-travel} sub-submodel (section \ref{buses-travel}).

\subsection{Sub-submodel: \codeTitle{cars-travel}} \label{cars-travel}
\begin{figure}[h]
    \centering
    \includegraphics[width=1\linewidth]{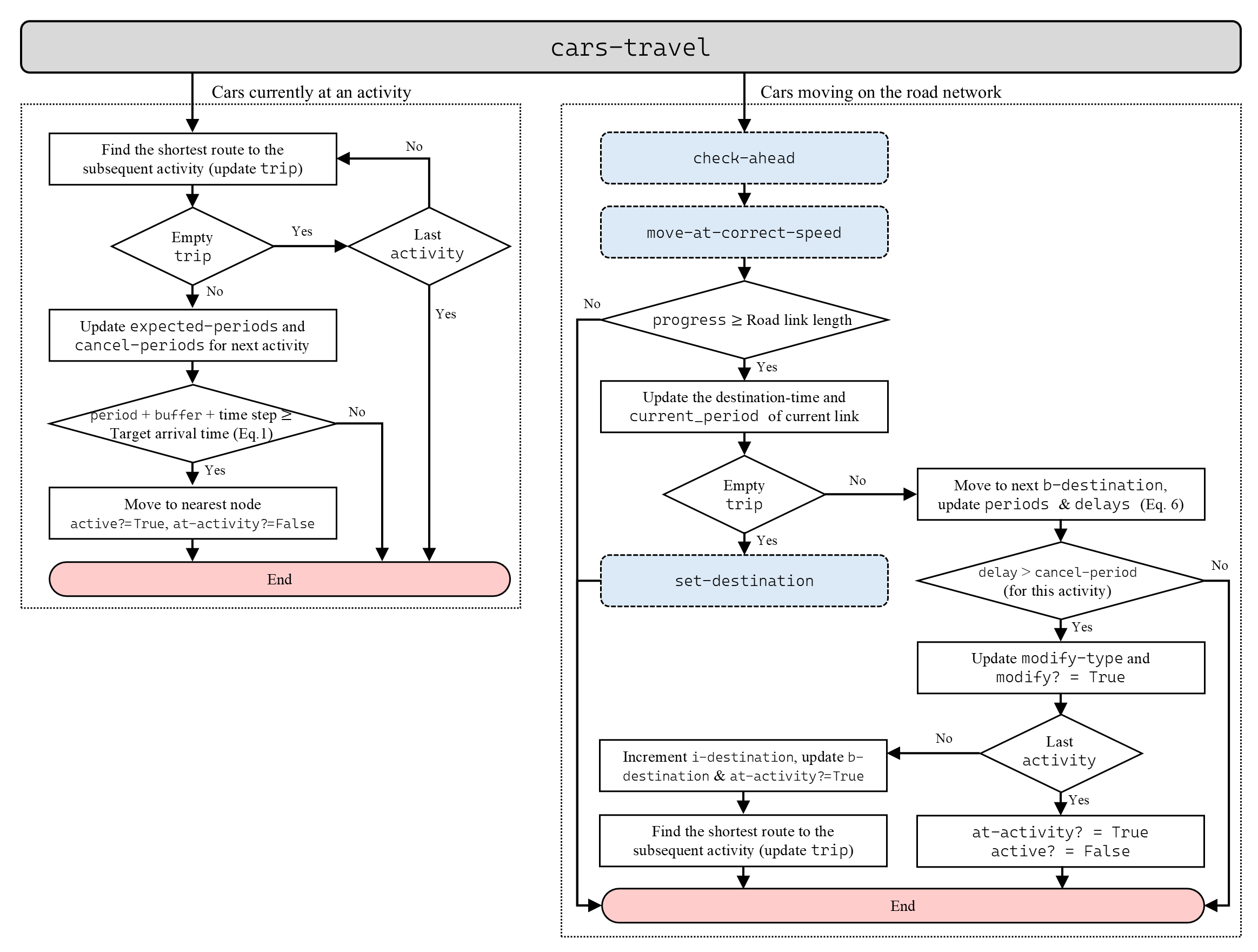}
    \caption{\code{cars-travel} sub-submodel logic flowchart}
    \label{fig:1}
\end{figure}

The \code{cars-travel} sub-submodel is executable for cars. Its flowchart logic is shown in Figure \ref{fig:1} and described hereafter.
First, the \code{cars-travel} sub-submodel addresses cars currently undergoing an activity (\code{at-activity? = True}) in its respective \code{b-destinations}.
For each of those cars, the submodel iterates through the subsequent activity in the \code{activity} state variable. It identifies the shortest route from the nearest node of the current car location to the nearest node of the respective activity building in the \code{b-destinations} list. This shortest route minimises the total \code{distance\_to\_destination} between the nodes, rather than trip times. The generates an ordered list of links (route) are assigned to the \code{trip} state variable of the car. If the \code{trip} includes no list, this implies the current and next activity has the same nearest node -- therefore, the car does not need to move. In this case, the submodel skips this activity, proceeds to the next one and repeats the same process of finding routes and defining trips. This is repeated until an activity with a trip including at least one road link is found, or until no subsequent activities exist. The outcome of this process is an assigned trip for the next activity of a car. 
The sub-submodel then checks if the car's trip does not include any links and if the car is at its last \code{b-destination}. In that case, the car is deemed as inactive (\code{active? = False}). This ensures that the car is no longer addressed in this sub-submodel or in \code{vehicles-move} submodel in the next time steps.
The sub-submodel then iterates through road links in the generated trip. For each road link, it accesses the \code{current\_period}. It then calculates the sum of the \code{current\_period} values of all the links in the \code{trip}. This replaces the \code{expected-periods} for the next activity, which implies the car considers the effect of current congestion patterns on trip time. The sub-submodel then calculates the period at which the car will cancel the trip for the next activity as per equation \ref{eq:3}. This value replaces the element in the \code{cancel-periods} variable at the respective next activity index. This ensures that the \code{cancel-periods} are updated based on the current congestion patters.
The sub-submodel then identifies whether the car should start moving to the next destination or not. It compares the target arrival time \(t\) to the next activity current time step (\(T\)), the \code{buffer-period} (\(b\)) and the current expected period (\(p\)) as per equation \ref{eq:1}. If the condition in equation \ref{eq:1} is satisfied, the submodel moves the car to the nearest node of its current location if no other cars are currently at that node, and it labels it as not at an activity (\code{at-activity?=False}). This assures that no unintended collisions occur at the beginning of the car's trip.

Second, the \code{cars-travel} sub-submodel addresses the cars currently not engaged in an activity (\code{at-activity? = False}). These cars are currently making a trip on the road network to their next activity. The sub-submodel initiates the \code{speed} of the car as its \code{top-speed}. It then initiates the \code{check-ahead} sub-submodel as described in section \ref{check-ahead}. This modifies the \code{speed} of the car to assure no collisions occur. Subsequently, the \code{move-at-correct-speed} sub-submodel is executed as described in section \ref{move-at-correct-speed}.
The sub-submodel then checks if the car reached end of its current road link in the \code{trip}. This is applied by calculating the length of the road link to the \code{progress} distance the car made from the origin node of the road link. If the \code{progress} is higher than length of the road link, then the car had reached the end of that link. In that case, the sub-submodel updates the \code{destination-time} of the car as the current time step. It then assigns the \code{current\_period} of the completed road network as shown in equation \ref{eq:5}.
\begin{equation} \label{eq:5}
    r_k = a_{i,n_1} - a_{i,n_2} 
\end{equation}
where \(r_k\) is the \code{current\_period} a car takes to move across the road link \(k\),
\(a_{i,n_1}\) is the \code{location-time} representing the time step at which a car \(i\) reached the start node \(n_1\) of the road network \(k\) and
\(a_{i,n_2}\) is the \code{destination-time} representing the time step at which a car \(i\) reached the end node \(n_2\) of the road network \(k\).

The sub-submodel then checks the \code{trip} of the car. If the \code{trip} includes links, meaning the car has to continue its trip, the sub-submodel \code{set-destination} is executed as described in section \ref{set-destination}. This modifies the \code{trip}, allocates the car's new road link start node (\code{location}) and end node (\code{destination)} and ensures any progress distance made beyond the previous road end node is not lost.
If the \code{trip} includes no links, meaning the car finished its trip, the car is moved to its current \code{b-destination}. The car adds the period it took to reach the activity to its \code{periods} list. It also calculates the delay as per equation \ref{eq:6} and adds it to the list \code{delays}. 
\begin{equation}\label{eq:6}
    y_{i,j} = T - a_{i,j}
\end{equation}
where \(y_{i,j}\) is the delay time steps for incurred by the car \(i\) to reach the activity destination \(j\) and 
\(a_{i,j}\) is the arrival time at the activity destination \(j\)

The car then checks if the period to taken to reach the activity is higher than its respective cancel period (in the \code{cancel-periods} state variables). If so, the car adds the modifies the respective element in the to its \code{modify-type} list to "cancel", and it flags it requests an activity modification (\code{modify? = true}). The car then checks if its current \code{b-destination} is the last building in its \code{b-destinations} list. If so, the car flags itself as at an activity (\code{at-activity = True}) and inactive \code{active? = false}. Otherwise, the car increments its \code{i-destination} to refer to the next activity index, updates its \code{b-destination} and flags itself as at an activity (\code{at-activity? = True}). The car then finds the shortest route to its next \code{b-destination} and assigns the sequence of road links to its \code{trip}.

\subsection{Sub-submodel: \codeTitle{buses-travel}} \label{buses-travel}
\begin{figure}[h]
    \centering
    \includegraphics[width=1\linewidth]{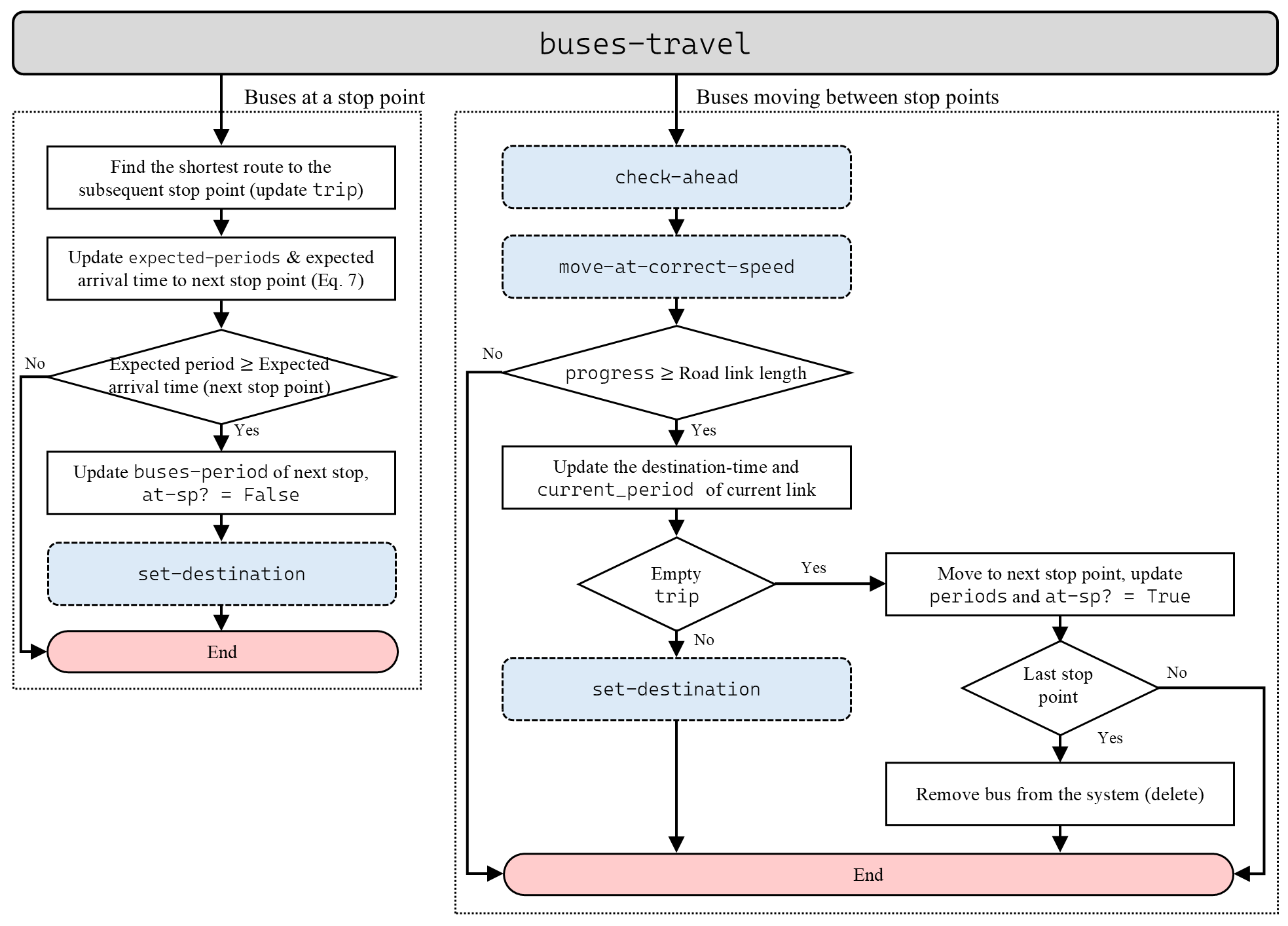}
    \caption{\code{buses-travel} sub-submodel logic flowchart}
    \label{fig:2}
\end{figure}

The \code{buses-travel} sub-submodel addresses buses, and it follows a similar logic to the \code{cars-travel} sub-submodel (section \ref{cars-travel}). The flowchart of the \code{buses-travel} sub-submodel is shown in Figure \ref{fig:2} and described hereafter.

First, the \code{buses-travel} sub-submodel addresses buses currently at a stop point (\code{at-sp? = True}). For each of those buses, the sub-submodel finds the shortest route from their current stop point (the bus's \code{location}) to the next stop point (the bus's next item in \code{sp-destinations}. The sub-submodel assigns the sequence of links of the shortest route's to the bus's \code{trip} state variable.  It then iterates through the road links in the \code{trip} to access each link's \code{current\_period}. This allows the sub-submodel to identify the total expected period for the trip and update the respective element in the bus's \code{expected-periods} list. The calculated expected period is then used to identify the expected arrival time of the bus in the next stop point as shown in equation \ref{eq:7}.
\begin{equation}\label{eq:7}
    a_{s+1} = T + p_{s\rightarrow s+1}
\end{equation}
where \(a_{s+1}\) is the arrival time at stop point \(s+1\) and
\(p_{s\rightarrow s+1}\) is the period taken to travel from stop point \(s\) to stop point \(s+1\).

The sub-submodel then checks the condition \(a_{s+1} \leq t_{s+1}\) -- where \(t_{s+1}\) is the target arrival time to the next stop point. If this condition is satisfied, the bus is expected to either arrive on-time \(a_{s+1} = t_{s+1}\) or late \(a_{s+1} > t_{s+1}\) to the next bus stop. In both cases, the \code{buses-period} of the next stop point is modified to reflect the expected arrival time. This assures that pedestrians plan their trip in case of a late bus arrival. The bus then executes the \code{set-destination} sub-submodel as described in section \ref{set-destination}. This manages the \code{trip} and allocates the start and end nodes of the first road link as the \code{location} and \code{destination} respectively.

Second, the \code{buses-travel} sub-submodel addresses buses currently making a trip from one stop point to another. It assigns the \code{speed} of the bus as its \code{top-speed}. The \code{check-ahead} sub-submodel is then executed as described in section \ref{check-ahead}. This updates the \code{speed} of the buses to avoid collisions. Following this, the \code{move-at-correct-speed} sub-submodel is applied as per section \ref{move-at-correct-speed}. The \code{buses-travel} sub-submodel then compares \code{progress} of the bus to the length of the road link it is currently travelling on. If the \code{progress} is higher, then the car has reached the end point of that road. Accordingly, the sub-submodel updates the \code{destination-time} of the bus (destination refers to the road link end node). It also assigns the \code{current\_period} of the road link as per equation \ref{eq:5}. This assures that cars and buses planning their next trips consider updated travel times reflecting congestion patterns. 
The \code{trip} is then checked. If it includes links, the sub-submodel \code{set-destination} is applied as per section \ref{set-destination}. This assures the bus moves to the next link in the \code{trip} and  modifies the \code{trip} accordingly. It also assures any \code{progress} made beyond the length of the previous road link is incurred in the next road link.
If the \code{trip} does not include include any links, the bus moves to its target \code{sp-destination} signalling an end to its current trip. The bus updates its \code{periods} to include the time period it took to reach the \code{sp-destination}.

The bus then checks if the reached stop point (\code{sp-destination}) is its last stop point (last item in \code{sp-destinations}). If the condition satisfied, the bus is removed from the simulation system. Otherwise, the bus increments its \code{i-destination} to refer to the next item in \code{sp-destinations}. It accordingly updates its \code{sp-destination}, and it also removes the previous stop point from the \code{remaining-sp-desitnaitons}. The bus flags it is currently at a stop point \code{at-sp? = True}. It also finds the shortest route to its next stop point from its current location. The sequence of road links are assigned to the \code{trip} state variable.

\subsection{Sub-submodel: \codeTitle{buses-start-SPs}} \label{buses-start-SPs}
The \code{buses-start-SPs} addresses the bus templates to create buses.
The sub-submodel accesses the stop point nodes from which buses should start this time step (i.e., nodes with  \code{0 $\in$ buses-period}). It extracts a list of these starting bus service numbers and accesses their respective bus templates. Accordingly, it updates the \code{buses-period} of each respective stop point node as the \code{frequency} of the respective bus's template. This assures that the next bus will be created at the correct time step.
The sub-submodel then creates buses which inherit the state variables of their respective bus templates. The bus template state variables values are shown in the \code{load-buses} sub-model (section \ref{load-buses}), and the input data to indicate the \code{service} number and \code{frequency} of the model are described in section \ref{Input data}.

\subsection{Sub-submodel: \codeTitle{check-ahead}} \label{check-ahead}
The \code{check-ahead} sub-submodel is applicable for both cars and buses (labelled as vehicles). The sub-submodel considers one vehicle at a time. 
First, it finds the set of vehicles that satisfy five conditions: (1) the vehicle is not at an activity (\code{activity = False}); (2) the \code{destination} node of the car or bus is the same as the \code{destination} of the considered vehicle; (3) the car or bus made more progress (\code{remaining-progress}) on the road link than the considered vehicle; (4) the direction the car or bus is facing is the same as that of the considered vehicle; and (3) the car or bus is within one model space unit distance from the considered vehicle. Out of all the cars and buses satisfying the aforementioned conditions, the closest one to the considered vehicle is selected. This is assigned as the nearest car or bus from the considered vehicle  travelling in its same direction.

Second, it addresses the \code{speed} of the considered vehicle (labelled \(i\)) based on the \code{speed} of and distance from the closest car or bus (labelled \(i'\)). If the closest car or bus is not moving (\code{speed = 0}), the considered vehicle also stops. If the closest car satisfies the condition \(D_{i\leftrightarrow i'} > 2.v_{i}\), the considered vehicle stops -- where \(D_{i\leftrightarrow i'}\) is the distance between \(i\) and \(i'\) and \(v_{i}\) is the speed of \(i\) per time step. If the closest satisfies the condition \(D_{i\leftrightarrow i'} > 3.v_{i}\), the considered vehicle checks the \code{speed} of the closest bus or car \(i'\). If \(v_i > v_{i'}\), the considers vehicle lowers its \code{speed} \(v_i\) to match that of the ahead vehicle \(v_{i'}\).

\subsection{Sub-submodel: \codeTitle{set-destination}} \label{set-destination}
The \code{set-destination} sub-submodel is executable for both cars and buses (vehicles). It is called once a vehicle reaches the end node of the road link it is currently moving through or when the vehicle is starting its trip. The sub-submodel extracts the first element of the \code{trip} list, which is a road link. It then defines the \code{location} as the road node where the vehicle is at, and it defines the \code{destination} as the other road node. It then removes the first element of the \code{trip} state variable. This assures the \code{trip} represents the remaining road links to reach the activity destination, excluding the current one.

\subsection{Sub-submodel: \codeTitle{move-at-correct-speed}} \label{move-at-correct-speed}
The \code{move-at-correct-speed} sub-submodel is applicable for both cars and buses. 
First, it assures the vehicle's \code{speed} is the lower of two values: the \code{speed\_limit} of the current road; and the \code{top-speed} of the vehicle. Second, it spatially moves the vehicle forward the distance covered at its \code{speed} per time step. This distance varies on changing the input \code{tick-time-scale} global state variable.
Third, it updates the \code{remaining-progress} of the car to reflect the distance between the vehicle and the \code{destination} road node after the vehicle moved.

\subsection{Submodel: \codeTitle{pedestrians-move}} \label{pedestrians-move}
The \code{pedestrians-move} sub-submodel manages pedestrians. If a pedestrian is active (\code{active? = True}), it calls the \code{pedestrians-travel} sub-submodel as described in section \ref{pedestrians-travel}.

\subsection{Sub-submodel: \codeTitle{pedestrians-travel}} \label{pedestrians-travel}
\begin{figure}[h]
    \centering
    \includegraphics[width=1\linewidth]{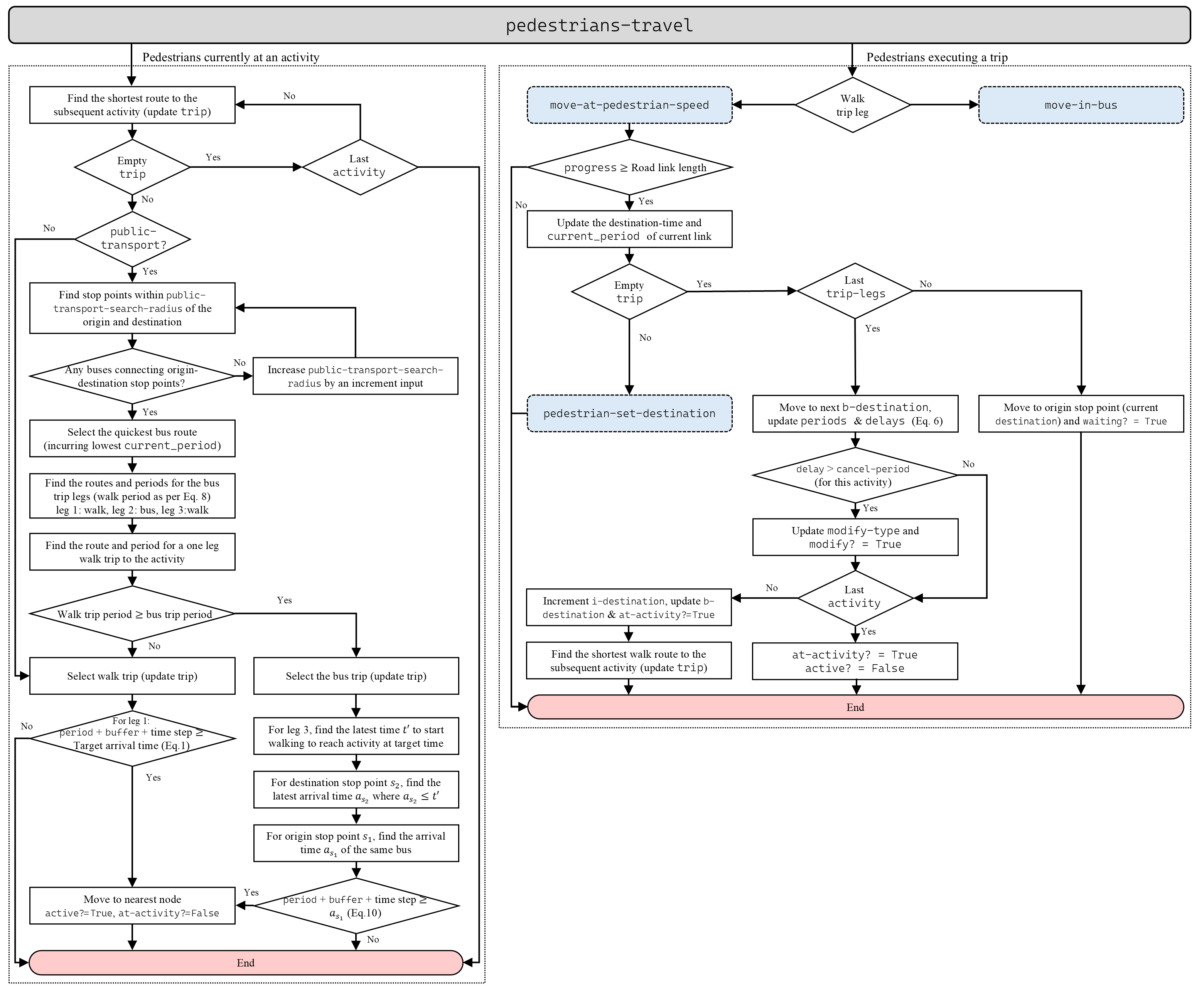}
    \caption{\code{pedestrians-travel} sub-submodel flowchart}
    \label{fig:3}
\end{figure}

The \code{pedestrians-travel} follows a similar logic to the \code{cars-travel} sub-submodel (section \ref{cars-travel}) for walk trips, and it deviates for bus trips. A flowchart of the logic of the \code{pedestrians-travel} sub-submodel is shown in Figure \ref{fig:3} and described hereafter.

First, the \code{pedestrians-travel} sub-submodel addresses pedestrians currently at an activity (\code{at-activity? = True}). For each of those pedestrians, the shortest route to the next activity in the \code{activity} list state variable is identified. The sequence of road links in that route are assigned to the \code{trip}. If the \code{trip} happens to not have any links, this implies that the next activity has the same nearest node on the road network as the current location of the pedestrian. Accordingly, the sub-submodel skips this activity and iterates to the subsequent one. This is repeated until a \code{trip} with at least one road link is found or until there are no subsequent activities. If it is the latter case, the pedestrian moves to its last building in the \code{b-destinations} list and it is flagged as inactive (\code{active? = False}). Inactive pedestrians are no longer addressed in this sub-submodel.

The \code{pedestrians-travel} sub-submodel then finds the optimum trip, including potential use of buses. If the pedestrian uses buses (\code{public-tranpsort? = true}), the sub-submodel finds two optimum trips: a trip including one bus and a trip including walking only. For the bus trip, the sub-submodel identifies the stop point surrounding the pedestrian's location (origin stop points) and surrounding the pedestrian's activity destination (destination stop points) within the \code{public-transport-search-radius}. The bus services linking the origin and destination stop points are then identified -- these are viable bus routes for the pedestrian's trip. If no viable bus routes are found, or if no origin or destination stop points are found in the first place, the radial search region is expanded by the \code{search-radius-increment-distance}. This process is repeated until at least at least one viable bus route is found. The sub-submodel then finds the quickest trip given the alternative viable bus routes. It subdivides the trip into three legs: (1) a walk leg to the origin stop point; (2) a bus trip from the origin stop point to the destination stop point; and (3) a walk leg to the activity destination. The walk legs are defined as the quickest route from each respective origin and destination. To find this route, the shortest route yielding the lowest \code{distance\_to\_destination} is defined. The expected travel period for that route is assured to be the lowest as pedestrians are not restricted to the \code{speed\_limit} of the road links. The expected period is calculated for each leg as shown in equation \ref{eq:8}.
\begin{equation}\label{eq:8}
    p_{i,m_1\rightarrow m_2} = \frac{\sum_{l\in L_{m_1\rightarrow m_2}}d_{l}}{v_i}
\end{equation}
where \(p_{i,m_1\rightarrow m_2}\) is the walk travel period from an origin \(m_1\) to a destination \(m_2\),
\(L_{m_1\rightarrow m_2}\) is the set of road links connecting \(m_1\) and \(m_2\) through the shortest route,
\(d_l\) is the length of the road link \(l\) in the set of road links \(L_{m_1\rightarrow m_2}\) and
\(v_i\) is the speed of the pedestrian \(i\).

The bus leg is defined as the quickest route from the origin stop point to the destination stop point. That is, the route that includes the sequence of road links incurring the lowest sum of \code{current\_period} state variables. For each three legged trip, the total time period to finish the three legs is calculated, and the trip with the lowest time period is selected as the optimum bus trip. Following this, the pedestrian identifies the shortest walk route from its location to the activity destination. That is, the route including the road networks incurring the lowest sum of \code{distance\_to\_destination}. The travel period across that walk trip is calculated as previously described in equation \ref{eq:8}. The sub-submodel then compares the two optimum full walk and bus trips, and it selects the one with the lowest travel period. This decision indicates whether the pedestrian will use buses or not. Accordingly, the \code{trip-legs} (list of road links per leg), \code{trip-legs-modes} ("walk" or "bus") and \code{trip-legs-expected-periods} (expected time needed to finish each leg) are updated. It should be noted that a bus trip element in the \code{trip-legs} is defined as a list indicating the origin stop point, the bus template for the used bus service and the destination stop point.

If the pedestrian does not use buses (\code{public-transport? = False}), the shortest route is identified as per the walk only trip for pedestrians considering buses. The trip parameters are also updated as previously mentioned.

The \code{pedestrians-travel} sub-submodel then checks whether each pedestrian should move from its location or not to arrive at the destination of its first trip leg. If the pedestrian has a full walk trip, it compares the \code{target-times} to the \code{expected-periods} and \code{buffer-period} as shown in equation \ref{eq:1}. If the condition in equation \ref{eq:1} is satisfied, the considered pedestrian moves to the first node in its first trip leg and executes the \code{pedestrians-set-destination} sub-submodel as described in section \ref{pedestrians-set-destination}.

 If the pedestrian is planning to use a bus, the sub-submodel traces its trip back from the last leg in the \code{trip-legs} state variable. For the last leg, the sub-submodel considers the latest time the pedestrian must start walking from the destination stop point to the activity as per equation \ref{eq:9}.
\begin{equation} \label{eq:9}
    t'_{i,s_2} = t_{i,j+1} - p_{i,s_2\rightarrow j+1} + b_{i,j+1} 
\end{equation}
where \(t'_{i,s_2}\) is the latest time to start moving from the destination stop point \(s_2\) to arrive to the next activity \(j+1\),
\(t_{i,j+1}\) is the target arrival time to the next activity \(j+1\),
\(p_{i,s_2\rightarrow j+1}\) is the time period taken to walk from the destination stop point \(s_2\) to the activity \(j+1\) and
\(b_{i,j+1}\) is the buffer period.

The sub-submodel then accesses the \code{buses-schedules} of the destination stop point for the selected bus service. It identifies the latest time satisfying the condition \(a_{s+2}\leq t'_{i,s_2}\) -- where \(a_{s+2}\) is the arrival time of the selected service at the destination stop point. Given \(a_{s_2}\), the sub-submodel traces the respective arrival time of the same bus at the origin stop point of the pedestrian \(a_{s_1}\) -- where \(s_1\) is the origin stop point. Accordingly, the arrival time to the destination of the first leg (i.e., the origin stop point \(s_1\)) is considered as \(a_{s_1}\). The sub-submodel considers a similar condition to equation \ref{eq:1} as shown in equation \ref{eq:10}.
\begin{equation} \label{eq:10}
    T + p_{i,j\rightarrow s_1} + b_{i,j+1} \geq a_{s_1}
\end{equation}
where \(p_{i,j\rightarrow s_1}\) is the time period taken to walk from the activity \(j\) to the origin stop point \(s_1\).

If the condition in equation \ref{eq:10} is satisfied, similar to the full walk trip pedestirans,it moves to the first node of its first trip leg and executes \code{pedestrians-set-destination} as per section \ref{pedestrians-set-destination}.

Second, the \code{pedestrians-travel} sub-submodel addresses pedestrians currently making a trip to an activity destination. These pedestrians fit into three categories: (1) walking, (2) on a bus; (3) and waiting for a bus. 
If the pedestrian is walking (current \code{trip-legs-modes = "walk"}), it executes the \code{move-at-pedestrians-speed} sub-submodel as described in section \ref{move-at-pedestrians-speed}.
If the pedestrian is on a bus or waiting for a bus (current \code{trip-legs-modes = "bus"}), it executes the \code{move-in-bus} sub-submodel as described in section \ref{move-in-bus}. This addresses the behaviours of pedestrians waiting for a bus (\code{waiting? = True}) and pedestrians on a bus (\code{on-bus? = True}).
If the pedestrian is waiting for a bus, it takes no action.

The sub-submodel then considers pedestrian who made more \code{progress} than the length of its current road links for the three aforementioned pedestrians categories.
If the pedestrian is walking and still has subsequent links in its \code{trip}, it executes the \code{pedestrians-set-destination} sub-submodel as described in section \ref{pedestrians-set-destination}. The pedestrian then executes the \code{move-at-pedestrians-speed} as described in section \ref{move-at-pedestrians-speed}. This assures the \code{trip} state variable is updated to reflect the remaining links in the current leg. It also assures the \code{progress} made beyond the length of the road link is not lost.
If the pedestrian is walking, has no subsequent links in its \code{trip} and is on its final trip leg, it implies it reached its activity destination. Accordingly, it moves to the respective building in its \code{b-destination} and updates its \code{location} accordingly. It also adds the time period it took to reach the activity to its \code{periods} list. It then calculates the delay, similar to cars, as indicated in equation \ref{eq:6}. The pedestrian then checks if the period incurred to reach the activity is higher than the respective cancel period in the \code{cancel-periods} state variable. If so, the pedestrian flags an activity modification by updating the respective \code{modify-type} item to "cancel" and setting \code{modify? = True}. The pedestrian then checks if it has no subsequent buildings in its \code{b-destinations} list. If so, the pedestrian indicates it is inactive \code{active? = False} and that it is at an activity \code{at-activity? = True}. Otherwise, the pedestrian finds the shortest route to the next item in the \code{b-destinations} list. This is adjusted in a later time step to find the optimum walk or bus trip.
If the pedestrian is walking, has no subsequent links in its \code{trip} and is not on its its final trip leg, this implies it reached its origin stop point. In this case, it moves to that stop point (in case it made more \code{progress} at its last link), and it flags it is waiting for a bus \code{waiting? = True}.
If the pedestrian is on a bus, has no subsequent links in its \code{trip} and is not on its final trip leg, this implies it reached its destination stop point. In this case, it updates its \code{trip} to reflect the next walk trip in its \code{trip-legs}. If the trip happens to include no links, this implies that the destination stop point is the closest node to the activity destination. In this case, the pedestrian updates its \code{location} and \code{destination} to that stop point -- this prepares the pedestrian to execute the \code{move-at-pedestrians-speed} in the next time step. If the walk trip includes links, the pedestrian executes the \code{pedestrians-set-destination} to assure it updates its \code{trip}, \code{location} (start node of the first trip link) and \code{destination} (end node of the first trip link).
If the pedestrian is on a bus, it takes no action as all the movement is managed in the previously executed \code{move-in-bus} sub-submodel.

\subsection{Sub-model: \codeTitle{pedestrians-set-destination}} \label{pedestrians-set-destination}
The \code{pedestrians-set-destination} sub-submodel is executable for pedestrians, and it follows a similar logic to the \code{set-destination} sub-submodel (section \ref{set-destination}). It is called if a pedestrian is starting a walk trip leg or has reached the end node of its current road link on a walk trip leg. The \code{pedestrians-set-destination} extracts the first road link in the \code{trip}. It then defines the \code{location} and \code{destination} as the start and end road link nodes from the pedestrian's perspective. It then updates the \code{trip} to not include that road link to ensure it represents the remaining links in the trip only. It should be noted that the \code{pedestrians-set-destination} sub-submodel is separated from the \code{set-destination} one to allow for flexibility with assigning different behaviours for pedestrians and vehicles in future expansions of MATraM.

\subsection{Sub-submodel: \codeTitle{move-at-pedestrians-speed}} \label{move-at-pedestrians-speed}
The \code{move-at-pedestrian-speed} sub-submodel is applicable for pedestrians. It follows a similar logic to the \code{move-at-correct-speed} (section \ref{move-at-correct-speed}) except that it does not consider road links \code{speed\_limit}. It spatially moves the pedestrian forward the distance covered at its maximum speed while considering the time step scale \code{tick-time-scale}. It also updates the \code{remaining-progress} of the pedestrian to reflect the remaining distance to the end node of the current road link.

\subsection{Sub-Submodel: \codeTitle{move-in-bus}} \label{move-in-bus}
The \code{move-in-bus} sub-submodel is applied for pedestrians waiting for a bus (\code{waiting? = True}) and on a bus (\code{on-bus? = True}).
If the pedestrian is waiting for a bus, it identifies the bus services that connects its origin stop point and destination stop point. It then checks if any of those services are currently at its origin stop point (i.e., its \code{location}). If it finds one, it increments the \code{capacity} of that bus by 1, and it creates a passenger link with the bus. This link ties the movement of the pedestrian spatially with that bus. The passenger then updates its \code{current-bus}, flags that it is on a bus (\code{on-bus? = False}) and flags it is no longer waiting for a bus (\code{waiting? = False}). It updates its \code{location} as the \code{current-bus}, and it updates its \code{destination} as its destination stop point. It also updates the \code{trip} to only include the destination stop point -- noting that the \code{trip} for a bus trip is initiated with the origin stop point, the bus template and the destination stop point.

If the pedestrian is on a bus, and the bus has reached the pedestrian's stop node destination, the passenger deletes the passenger link with the bus. It flags it is no longer on a bus (\code{on-bus? = False}) and it is no longer waiting for a bus (\code{waiting? = False}). It downgrades the bus \code{capacity} by 1 and updates its \code{current-bus} as none.

\section{Discussion} \label{Discussion}

The development of MATraM highlights both the potential and the current limitations of activity-based transport modelling. By introducing an agent-based framework capable of structuring daily activities while explicitly simulating mobility, this work seeks to overcome a well-recognised rigidity in existing approaches. Conventional activity-based models typically rely on fixed daily schedules, which constrains their ability to reflect behavioural adaptation. This rigidity has important implications. It complicates model calibration, as uncertainties embedded in initially assigned activity schedules are effectively ignored, and it limits the capacity of models to explore how individuals might respond to changes in the urban environment, such as policy interventions or infrastructure disruptions.

MATraM addresses these challenges through a flexible mechanism for identifying and responding to sub-optimal travel conditions. By enabling the flagging of activity modifications when trips deviate from expected performance, the model introduces a dynamic feedback between mobility outcomes and activity scheduling. This represents an important conceptual shift, as it allows behaviour to evolve endogenously rather than being imposed ex ante. As a result, MATraM is better positioned to capture the interplay between individual decision-making and system-level dynamics, particularly under conditions of uncertainty or change.

A key strength of the framework lies in its transferability. The parameters governing activity modification are not restricted to travel time alone but can be extended to incorporate alternative behavioural drivers. For example, frameworks such as OASIS \citep{Pougala2023} provide complementary parameterisations that could be embedded within MATraM to reflect broader activity generation and decision-making processes. While these extensions will require further empirical work to define appropriate thresholds and probabilities for behavioural change, the modular design of MATraM enables such integration, underscoring its value as a flexible modelling platform.

There remain several avenues for future research. Enhancing the representation of traffic dynamics, including lane-level interactions, would improve the realism of simulated mobility. Expanding the model to include additional transport modes, such as cycling, and incorporating mode-specific routing constraints would further broaden its applicability. In addition, closer integration with emerging activity modification frameworks, including recent GeoAI-based approaches \citep{Gamal2026} offers a promising direction for strengthening the behavioural realism of the model. Collectively, these developments would support the evolution of MATraM into a more comprehensive tool for analysing transport systems and their response to interventions.
\newpage

\newpage
\bibliographystyle{unsrtnat}

\bibliography{references}

\end{document}